\documentclass{aa}
\usepackage{graphicx}
\usepackage[mathscr]{eucal}
\begin{document}

\title{A Nulling Wide Field Imager for Exoplanets Detection and General Astrophysics}
       
\author{Olivier Guyon \inst{1,2} and Fran\c{c}ois Roddier}

\offprints{guyon@ifa.hawaii.edu}

\institute{Institute for Astronomy, University of Hawaii, 640 N. A'ohoku Place, Hilo, HI 96720 \and Universit\'e Pierre et Marie Curie, Paris 6, 4 Place Jussieu, 75005 Paris, France.}

\date{Received / Accepted }

\abstract{
We present a solution to obtain a high-resolution image of a wide field with the central source removed by destructive interference. The wide-field image is created by aperture synthesis with a rotating sparse array of telescopes in space. Nulling of the central source is achieved using a phase-mask coronagraph. The full (u,v) plane coverage delivered by the 60m, six 3-meter telescope array is particularly well-suited for the detection and characterization of exoplanets in the infrared (DARWIN and Terrestrial Planet Finder (TPF) missions) as well as for other generic science observations. Detection (S/N=10) of an Earth-like planet is achieved in less than 10 hours with a $1 \mu m$ bandwidth at $10 \mu m$.
}

\titlerunning{A Nulling wide-field imager for ...}
\authorrunning{Olivier Guyon \& Fran\c{c}ois Roddier}
\maketitle

\section{Introduction}
As seen from 10 parsec ($\:pc$) away, the Earth is 0.1 arc-second ($''$) away from the Sun and $10^{9}$ times fainter at visible wavelengths. At $10\: \mu m$, the light intensity ratio is only $10^{6}$, and a $20 \:m$ minimum baseline is needed to resolve the system. The lower Planet/Star intensity ratio as well as the likely presence of biomarkers ($CO_2,O_3,H_2O,CH_4$) in the atmospheric spectrum from $5\: \mu m$ to $15\: \mu m$ led us to choose this wavelength range for our study. A space interferometer with nulling capabilities is well suited for this work because of the large baseline needed to resolve the system and the high luminosity ratio between the planet and the star.

Bracewell (1978) proposed to combine the beams from 2 telescopes to form interferometric fringes with the star centered on a dark fringe (optical path-length difference of half a wavelength between the 2 light beams). The on-axis star's light would then be nulled, and by rotating the 2-telescope interferometer around its optical axis, the light from a companion would be temporally modulated. The same concept has more recently been generalized to several telescopes to obtain a deeper null and a better (u,v) coverage (\cite{ang97,men97}). A monopixel detector records a light intensity, which is the integral of the product of the light distribution on the sky and a ``transmission map'' given by the geometry of the array. As the interferometer rotates, the residual light of the on-axis star remains constant and an off-axis companion is revealed as a temporal modulation of the recorded signal. The imaging capabilities of these concepts are very limited (small field of view and very partial (u,v) plane coverage), increasing the risk of confusion between an exoplanet and an anisotropy of the exozodiacal cloud. The detection by light modulation on a monopixel detector suffers from high sensitivity to temporal variations of both the thermal background intensity and the light leakage from the nulled central star.
In this paper, we present a solution to image (with a full (u,v) plane coverage up to the cutoff frequency) a field with the central source removed by destructive interference. The wide field of view and good (u,v) plane coverage limit the risk of confusion with background sources and exozodiacal anisotropies.
We demonstrate the importance of good (u,v) coverage and show how to achieve it in \S2. The pupil densification and pupil redilution techniques are presented in \S3: these techniques make possible the use of a nulling coronagraph on a sparse array of telescopes. The overall behavior of the concept is studied in \S4. In \S5 and \S6, we explore the capabilities of this concept to image Earth-like planets and study other astrophysical objects.

\section{(u,v) plane coverage}

\subsection{Importance of good (u,v) plane coverage}
Very partial (u,v) plane coverage is sufficient for observation of objects for which a simple model exists. Ground-based interferometers are routinely measuring stellar diameters using 2 telescope-interferometers. By changing the projected baseline (by physically moving the telescopes or taking advantage of Earth's rotation), more points are accessible in the (u,v) plane and limb darkening can be measured (\cite{haj98}). Unfortunately, no simple model exists for a planetary system, and exoplanet imaging missions like TPF (\cite{bei98}) and Darwin (\cite{leg96}) need good (u,v) plane sampling before detecting with certainty an exoplanet. A planetary system is likely to have several planets embedded in a zodiacal cloud. As seen face-on, a planetary system with an Earth-like planet 1 Astronomical Unit ($AU$) from a Sun-like star in a 1 Zodi (unit of dust density in a zodiacal cloud, normalized to the solar system zodiacal cloud) zodiacal cloud is as bright as a $0.3\:AU$ by $0.3\:AU$ ``pixel'' of exozodiacal light next to it.

The presence of massive bodies in the zodiacal cloud will perturb the orbits of dust particles and produce an elongated concentration of zodiacal dust: in the solar system, the Earth is responsible for a 10\% variation of the surface brightness of the zodiacal dust as seen from Earth (\cite{rea95,der94}). Because in some cases these clumps can be brighter than the planet itself (\cite{lio97}), it has been proposed to infer the existence of a planet from observation of the exozodiacal cloud. Recent direct ground-based observation of the dust stream responsible for the Leonid meteor shower (\cite{nak00}) and IRAS/ISO observations of cometary trails (\cite{syk92,kre93,jen97}) demonstrate the long-lasting effect of cometary debris. In our solar system (seen face-on), the light intensity from dust concentrations in the zodiacal emission (excluding the smooth component of the zodiacal cloud), when integrated over a scale of a tenth of an $AU$, is less than a percent of the Earth's light intensity. However, in systems with a higher zodiacal component or no massive planet outside of the Earth's orbit, dust concentrations could be as luminous as an Earth-like planet. This becomes even more critical if the system is seen edge-on, since the surface-brightness of planet-induced arcs of dust would be amplified when seen tangentially.

The TPF and Darwin missions propose to image stars at a distance up to $20\:pc$ to look for planets. Being able to rule out arcs of exozodiacal light excess and background sources will make the detection of planets more reliable and, to some extent, reduce the need for repetitive observations of the same object to confirm a detection. A large field of view is also essential for characterization of solar systems from 5 to $20\:pc$. While imaging a Sun-Earth system at $20\: pc$ requires at least a 50 milli-arc-second ($\:mas$) resolution (separation corresponds to 1 resolution elements or more), the angular separation between a Sun-Jupiter system at $5\: pc$ is $1\:"$. A wide field of view requires a well-filled (u,v) plane, which also allows observations of various astrophysical sources for which no reliable model exists. 

\subsection{Obtaining full (u,v) plane coverage with aperture synthesis}
Several technical constraints limit the achievable (u,v) plane coverage: number of telescopes, telescope diameter and array geometry. A large number of telescopes or a larger diameter of individual telescopes would increase the cost and complexity of the mission. As the number of telescopes increases and their diameter decreases, the internal metrology of the array becomes more complex and less photons per aperture are available, making precise high-speed fringe tracking more difficult. We have chosen to work at $10\:\mu m$, and to reach the resolution needed to separate a Sun-Earth system at $20\:pc$ at this wavelength, a $60\:m$ baseline is needed ($34\:mas$ resolution at $10\:\mu m$). To further limit the number and size of telescopes, we have chosen to use rotation of the whole array to fill the (u,v) plane. Rotation of the array is the easiest way to change the set of (u,v) points measured, since it does not require a change in distances or relative angles between apertures (a rigid array can be used) and there is no need for large optical pathlength differences corrections.
Rotation of a rigid interferometric array greatly improves its (u,v) plane coverage. By adopting an optimal configuration, it is possible to obtain a full (u,v) plane coverage through rotation of an array of small telescopes. Guyon and Roddier (2001) have explored this problem for arrays of up to 10 identical telescopes and, for a given baseline, computed the configurations that allow full (u,v) plane coverages with the smallest possible telescope diameters.
Table 1 lists, for a given number of telescopes, the minimal telescope diameter required to fill the (u,v) plane up to the resolution limit of the array ($60\:m$).
\begin{table}[h]
{\small
\begin{tabular}{|c|c|c|c|c|c|c|c|}
\hline
N & 4 & 5 & 6 & 7 & 8 & 9 & 10\\ 
\hline
D(m) & 4.62 & 2.90 & 1.98 & 1.43 & 1.11 & 0.91 & 0.76\\
\hline
\end{tabular}
}
\caption{
Minimum telescope diameter (D) to fully cover the (u,v) plane up to the maximum frequency for a 60m array of N telescopes.  Adapted from (\cite{guy01}). \label{tbl-1}}
\end{table}

The array configuration for each of those options is given by Guyon \& Roddier (2001). From this table, six $2\:m$ telescopes can fully cover the (u,v) plane up to the maximum frequency of a 60m array. In this work, whenever numerical values are given, we choose to adopt the 6-telescope configuration, with $d=3\:m$ (diameter of each aperture of the interferometer) to obtain a good SNR at all spatial frequencies. The array geometry is shown in fig. 1 (upper left). In this paper, $N$ is the number of apertures, $d$ the diameter of each aperture and $\lambda$ is the wavelength.

\subsection{Tradeoff between (u,v) plane coverage and planet detection sensitivity}
Obtaining a complete (u,v) plane coverage with a limited number of small telescopes requires a non-periodic array. The Point Spread Function (PSF) obtained from a non-redundant array is poorly contrasted whereas a highly redundant array yields very contrasted series of diffraction peaks (Fig. 1). For planet detection, our concept uses both nulling (canceling most of the starlight photons) and imaging (separating the photons of the planet from the photons of the star on the detector). The lower contrast of the diffraction peaks reduces the signal to noise ratio (S/N) for detection of exoplanets because the spatial separation of the planet light and the star light is less efficient. We give some estimate of this decrease in S/N in Appendix B.
\begin{figure}
\centering
\includegraphics[width=8cm]{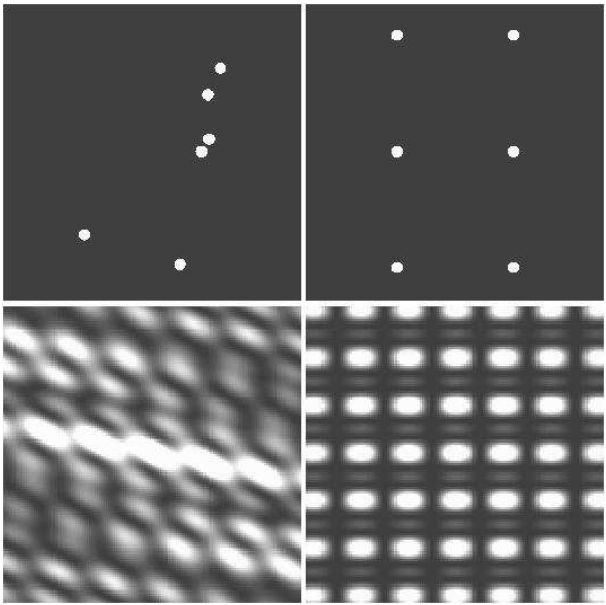}
\caption{Array configurations (top) and corresponding PSFs (bottom) for a non-redundant (left) and a redundant (right) array geometry. The diffraction peaks in the PSF of the redundant array are more contrasted. The non-redundant six telescope array on the top left is optimized for full (u,v) plane coverage when combined with rotational aperture synthesis.}
\end{figure} 

The efficiency of the (u,v) plane coverage could be sacrificed to gain S/N for planet detection at the expense of a reduced field of view and lower imaging capabilities (increased chances of confusion). The poorer imaging capabilities translate into a lower confidence in the detection, requiring multiple observations (longer exposure time). It is hard to estimate the overhead of a poorer (u,v) plane coverage because it is very dependent upon the structure of the object observed. For observations of complex astrophysical sources (galaxies, star forming regions) for which nulling is not critical, there is no such tradeoff and the coverage of the (u,v) plane should be optimized regardless of the instantaneous PSF contrast.

The nulling performance of our concept is independent of the array geometry. It is therefore possible to adopt a redundant configuration without any change in the concept. A reconfigurable array is also a possible solution for choosing the best array geometry for each type of observation.

In this work, we have adopted a nonredundant configuration which obtimizes the (u,v) plane coverage. This makes our concept a true imager, with excellent scientific capabilities for projects others than exoplanet detections. The risk of confusion between exoplanets, exozodiacal structures and background sources is also very low. Alternatively, the redundant array shown in Fig. 1 (top right) might also be considered for exoplanet detection and characterization: the better contrasted PSF allows this array to reach the same S/N ratio with exposure times 3 to 4 times shorter (see Appendix B). Although such a choice would be risky for observation of complex targets, this array would be most efficient on systems with a small number of planets and little exozodiacal structure.

\subsection{The image reconstruction algorithm}
A series of ``snapshot'' images is acquired with the array, with a direct recombinaition of the beams at the focal plane (Fizeau imaging). Each snapshot is acquired at a given rotation angle of the array. An example of aperture synthesis using this array, with rotation, is given in Guyon \& Roddier (2001). An important step of the reconstruction is the Fourier filtering of each snapshot frame to reduce the photon noise : Fourier components that are not sampled by the current configuration of the array are set to zero for each frame.

The rotation of the array can be continuous, provided that, if the snapshot exposure times are long (the array rotates significantly during one snapshot), the focal plane detector is counter-rotating in order to maintain a fixed orientation relative to the sky. Thanks to the linearity of the Fourier transform, each snapshot then samples a set of Fourier-space domains which are thick arcs rather than spots : the position of the objects on the detector is constant during an exposure, but the changing (rotating) diffraction patterns are integrated through the exposure.
Each snapshot is equivalent to an image acquired by a 60m telescope with mask in the pupil plane having six $3\:m$ diameter holes. Fizeau imaging is preferred to the classical fringe amplitude and phase measurements (one fringe per pair of telescopes) because collapsing the beam of each telescope into a coherent wave would limit the total field of view to $\lambda/d$ ($0.7\:"$ from edge to edge) which, in this example, is too small to image a Sun-Jupiter system at $10\:pc$.

\section{Pupil densification and nulling coronagraphy}

\subsection{Introduction to pupil densification}
\begin{figure}
\centering
\includegraphics[width=9cm]{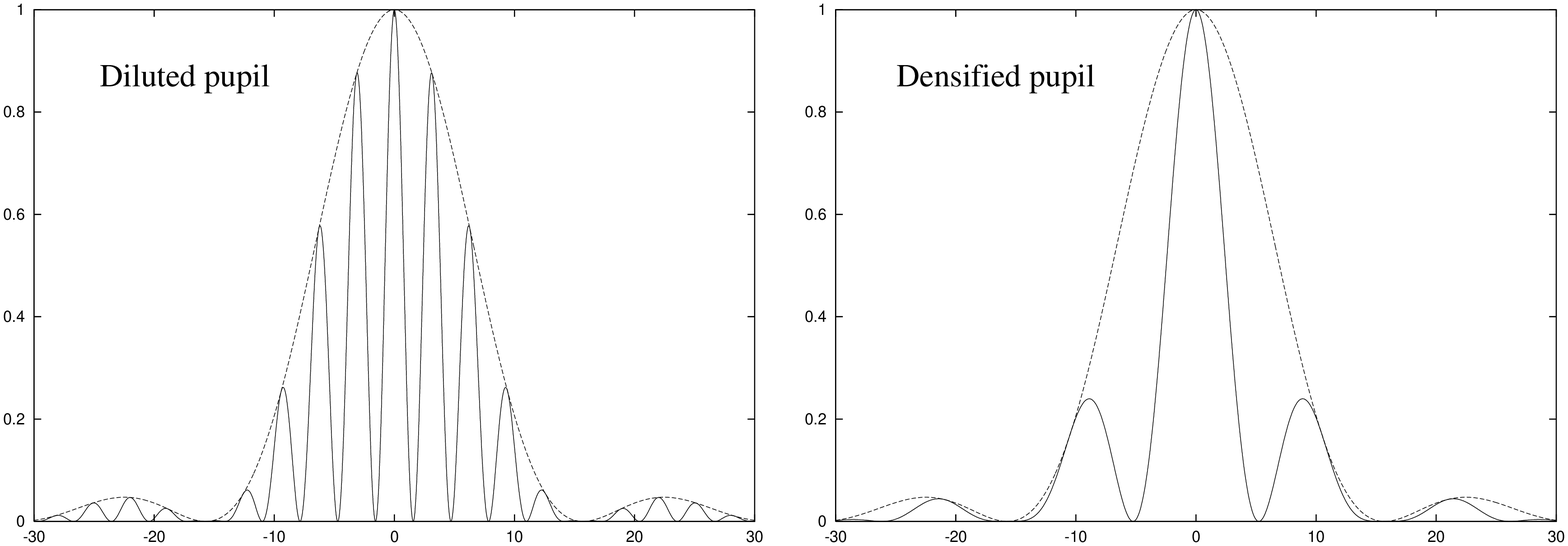}
\caption{Cut through a diluted pupil's PSF (top) and a densified pupil's PSF (bottom). A 2 apertures interferometer was simulated to generate the 2 plots. The envelope (PSF of a single aperture) is over-plotted on the PSF in each case. Because the distance between the 2 apertures is smaller when the pupil is densified, the fringe period is larger.}
\end{figure} 
\begin{figure}
\centering
\includegraphics[width=6cm]{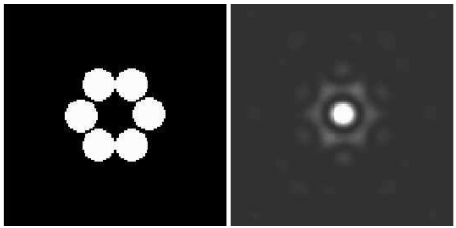}
\caption{Pupil (left) and PSF (right) in the densified pupil configuration for a 6 telescope array.}
\end{figure} 
The PSF of an interferometer (of baseline $B$ and sub-apertures diameter $d$) in Fizeau imaging is characterized by two quantities :
\begin{itemize}
\item The size of the central diffraction peak ($\lambda/B$)
\item The size of the PSF's envelope in which the diffraction peaks lie ($\lambda/d$).
\end{itemize}
Pupil densification increases the pupil filling factor by bringing the sub-pupils closer to each other, therefore artificially increasing $d/B$ and reducing the number of diffraction peaks in the PSF. It ``magnifies'' the diffraction pattern inside the envelope (fig. 2). When the pupil is fully densified, only one bright diffraction peak is inside the PSF's envelope. While the PSF is field-invariant in the diluted pupil scheme, it is not in the densified pupil scheme. Pupil densification of a sparse array of apertures creates a PSF close to a single-aperture telescope's PSF (\cite{lab96,ped00}) which allows a phase mask coronagraph to work efficiently. 
Figure 3 shows the densified pupil and corresponding on-axis PSF and is to be compared with the PSF in the diluted pupil configuration (Fig. 1, bottom left). In the diluted pupil configuration, the PSF has many bright secondary peaks while in the densified pupil configuration, most of the light is concentrated in the central peak.

\subsection{Requirements for coronagraphy and wide field of view imaging}
Because of the non-redundancy of the array (good (u,v) coverage), the diffraction peaks of the PSF in the diluted pupil scheme tend to break into multiple fringes, with lower contrast, away from the center of the PSF (Fig. 1). If a coronagraph were to be used in this focal plane, it would have to ``mask'' each of those peaks. Because of the ``fading'' of those peaks into multiple low-contrast fringes, this ``masking'' would also mask most of the light of any point source in the field of view. Another problem is the wavelength-dependence of the PSF. Diffraction peaks of the PSF are spectrally dispersed in a direction pointing to the center of the PSF. The combination of these two effects makes it impossible to achieve a good nulling of the star without also masking the planet.

A solution to the first problem would be to make the array redundant (regular spacing between apertures) to form high-contrast diffraction peaks across the PSF. The wavelength effect can be solved by using a wavelength-dependent magnification device to cancel out the wavelength-dependent scale of the PSF (\cite{wyn79,rod80}). Another solution to the wavelength problem is to use a 1D array to free up one dimension in the focal plane: this dimension can then be used to spectrally disperse the PSF on a coronagraphic device which follows the changing scale of the PSF on this axis. This is the solution that Aime et al. (2001) have explored with a linear periodic array. With a non-redundant, (u,v) coverage-optimized, array, pupil densification is needed to concentrate the light of the central source into a single high-contrast diffraction peak on which coronagraphy can be performed: the entrance pupil of the array is densified into a tight configuration before coronagraphy is applied.

Formation of a wide field of view ($FOV$) image ($FOV > \lambda/d$) requires Fizeau-mode imaging with a non-densified exit pupil. The number of resolution elements in the ``clean field of view'' of the densified pupil is too small to recover the information needed to create a wide field of view image using rotational aperture synthesis. This is why, after the coronagraphic device, the pupil needs to be rediluted into its original configuration (entrance pupil) before Fizeau combinaition of the beams on the focal plane detector (Fig. 4). Table 2 gives the Field of view and the number of diffraction peaks in each pupil configuration.
\begin{table}[h]
{\small
\begin{tabular}{|c|c|c|}
\hline
& Densified & Diluted \\
& Pupil & Pupil \\
\hline
Field of & Useful FOV & No\\
View & $\sqrt{N} \times \frac{\lambda}{B}$ & limitation\\
\hline
Number of & \raisebox{-1.5ex}[0pt]{1} & \raisebox{-1.5ex}[0pt]{$(\frac{B}{d\sqrt{N}})^2$} \\
diffraction peaks & & \\
\hline
\end{tabular}
}
\caption{
Field of view and number of diffraction peaks of the PSF in the densified pupil and diluted pupil schemes. $N$ is the number of apertures, $B$ the baseline and $d$ the diameter of individual apertures. \label{tbl-2}}
\end{table}

\begin{figure}
\centering
\includegraphics[width=6cm,angle=-90]{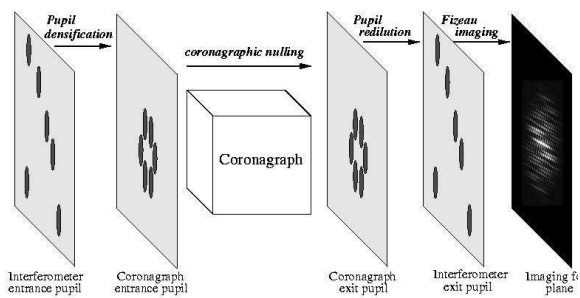}
\caption{The entrance pupil of the interferometer is densified before the coronagraph. The densified exit pupil of the coronagraph is rediluted before imaging on the detector array in order to maintain a wide field of view.}
\end{figure} 

\subsection{Sub-Aperture configuration in the densified pupil}
Coronagraphy works best when the densified pupil is tight (high filling factor). Such a tight configuration cannot be obtained if the relative positions of the sub-pupils' centers are kept constant: the pupil needs to be rearranged for an efficient densification. This seems to violate a common law of image formation through interferometers which states that the relative position of the sub-pupils of an interferometer should not be modified before image formation. In fact, this rule is not violated because the final image will be created after redilution of the array into its original configuration.

However, this means that the light complex amplitude at the coronagraphic focal plane does not follow the simple laws of the formation of an image. For an on-axis point source, the wavefront phase is constant on all sub-apertures and across them. This flat wavefront in the densified pupil will yield a sharp PSF in the coronagraphic focal plane. If the point source is not on-axis, wavefront slopes are present on each sub-aperture and large phase shifts exist between sub-apertures. In the entrance pupil, those wavefronts are all part of a large tilted wavefront, and the tilt corresponds to a translation of the PSF. In the densified pupil, the slopes and phase shifts cannot be obtained by multiplication of a tilted wavefront by the densified pupil function (1 inside the pupil and 0 outside): the PSF is not translation-invariant. For some positions of the point source, the PSF in the densified pupil case will be broken into several low-contrast diffraction peaks, while for others, the set of fringes will coherently add up into a single diffraction peak (Fig. 5).
\begin{figure}
\centering
\includegraphics[width=9cm]{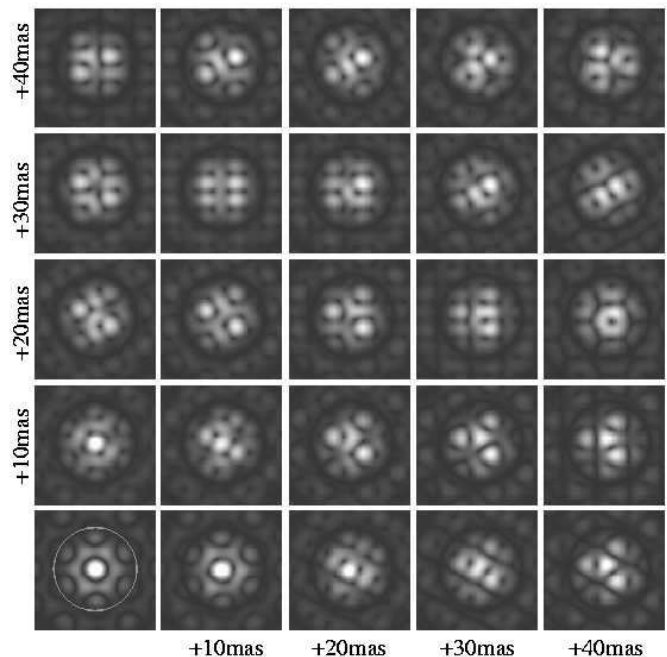}
\caption{PSFs in the coronagraph's focal plane (after pupil densification) for different positions on the sky. The lower left PSF corresponds to an on-axis point source.  In this grid, the point source position in the sky is incrementally increased by 10mas steps along two orthogonal axis. The white circle plotted on the on-axis PSFs indicates the ``clean field of view'' in the densified pupil scheme.}
\end{figure} 
For an on-axis point source, the coronagraph will ``mask'' the light of the source because a clear diffraction peak is centered on the optical axis.

\subsection{Nulling coronagraphy}
Pupil densification reduces the useful field of view in the coronagraph's focal plane to about $\sqrt{N} \times \frac{\lambda}{B}$ times $\sqrt{N} \times \frac{\lambda}{B}$. With a small number of telescopes ($N < 10$), this field of view is only a few times bigger than the diffraction peak of the PSF. There is therefore a very strong constraint on the size of the mask to use in this concept: a coronagraph whose mask would ``block'' an area equal to or larger than $\sqrt{N} \times \frac{\lambda}{B}$ times $\sqrt{N} \times \frac{\lambda}{B}$ in the coronagraph's focal plane would in fact ``block'' the light of any source closer than $\frac{\lambda}{d}$ to the optical axis. Figure 6 illustrates this effect by showing the transmission map of the concept on the sky when a phase mask coronagraph is used: the distance between consecutive nulls is comparable to the width of each null. If $N < 10$, a Lyot coronagraph shouldn't be used since a Lyot coronagraph, even with an apodized mask, cannot reach good null performance (better that $10^3$) with masks smaller than $2 \times \frac{\lambda}{d}$. The phase mask (\cite{rod97,guy99}), however, offers a very attractive solution because of its very small size (less than half the diameter of the first dark Airy ring for a circular aperture) and its theoretical total extinction for an on-axis point source. The four quadrants phase mask (\cite{rou00,ria01}) offers similar nulling performance but would yield a lower overall transmission due to partial extinction on the two perpendicular axes of the phase shifts, and is sensitive to the shape of the densified pupil (the four quadrants phase mask would not work on the densified pupil geometry we have adopted in this work). The phase mask coronagraph can work efficiently with almost any pupil shape, provided that the pupil is not too sparse, but the apodization mask required in the entrance pupil of the coronagraph (Appendix A) is different for different pupil shapes: if a single apodization mask is used, the densified pupil shape cannot be modified. However, the four quadrants phase mask does not suffer from the wavelength-dependant PSF scale. For $N<10$, the lower overall transmission of the the four quadrants coronagraph seriously affects the sensitivity of the study. However, for large values of $N$, this becomes a less serious problem, and the advantages of the four quadrants phase mask (no need for a pupil apodization mask, insensitivity to wavelength-dependant PSF scale) could make it a more attractive solution than the phase mask.

Therefore, we have chosen to use the phase mask coronagraph with an apodized densified pupil. This technique offers a total extinction of an on-axis monochromatic point source. Details of the apodization technique as well as performance and sensitivity of the phase mask to various errors are given in appendix A. The results of this detailed study of the phase mask coronagraph technique are used to estimate the sensitivity of the concept to point-source detection (Appendix B).

\section{Imaging with the phase mask coronagraph applied on a sparse array of telescopes}

\subsection{PSF characteristics}
Introducing a coronagraph after densification of the pupil (and before redilution) removes the invariance by translation of the PSF in the final focal plane (after pupil redilution). The light of an on-axis point source is stopped by the coronagraphic assembly while for other point source positions, most of the incoming light is detected in the final focal plane. This field-dependent modulation of the transmission (Fig. 6) is accompanied by a change of the PSF structure (Fig. 7). 

\begin{figure}
\centering
\includegraphics[width=7cm]{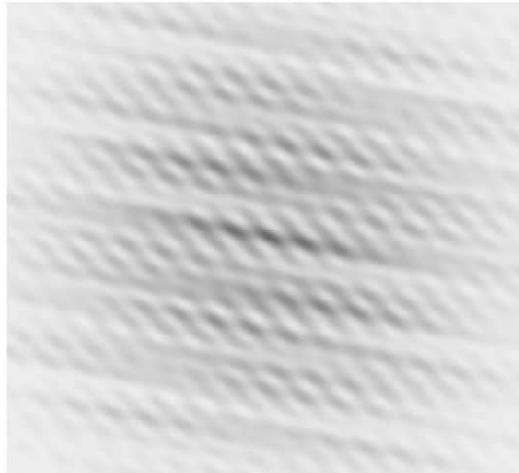}
\caption{Transmission map of the interferometer on the sky. The scale is linear from 0 (black, no transmission of the source's light) to 1 (white, total transmission). The image is $1.024\:''$ by $1.024\:''$.}
\end{figure} 

\begin{figure}
\centering
\includegraphics[width=9cm]{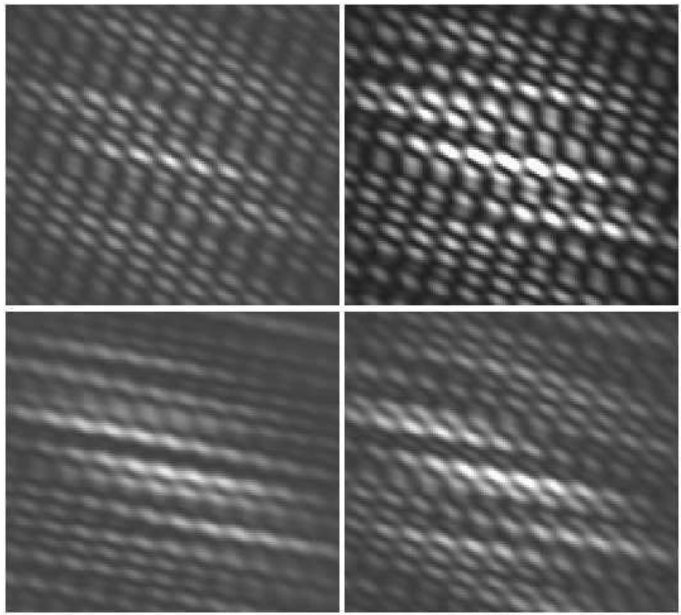}
\caption{PSFs of the concept for 4 different positions on the sky. Each frame is $1.6\:''$ by $1.6\:''$. Positions on the sky, in $mas$, relative to the optical axis, are (+10,+10) (upper left), (+20,+20) (upper right), (+30,+30) (lower left) and (+40,+40) (lower right).}
\end{figure} 

\subsection{Array rotation and residual star light removal}

The image of a star (disk of uniform brightness) on the focal plane is invariant with the rotation of the array (the detector is corotating with the array). The image of an off-axis point source (planet), however, is rotating around the optical axis and changing in structure (effects introduced by the coronagraph, see \S4.1). It is hence possible to decouple the static image of the star (and other rotationally symmetric light distributions) from the changing image of planets (and other non-rotationally symmetric light distributions). For example, the median of all the frames gives a very good approximation of the residual star light and can then be subtracted from each individual frame. When a high number of snapshots have been acquired at various position angles of the interferometer, the median of all snapshots yields a very reliable image of the rotationally symmetric component of the observed object (Star + rotationally symmetric component of the exozodiacal cloud). For detection of exoplanets (or other non-rotationally symmetric structures), this self-calibrating technique is expected to be significantly more reliable than observation of reference stars. Thanks to this technique, any noise which is not correlated with the position angle of the array (such as ``speckle'' noise) will be efficiently decoupled from the planet's signal and will average out efficiently.

Our simulations show that the distribution of the residual starlight on the detector has little dependence with the angular size of the star (provided that the star is not resolved). It can be approximated by a reference distribution multiplied by a coefficient which increases with the angular diameter and surface brightness of the star. This property allows precise modeling of the residual starlight in the focal plane.
Thanks to the very large number of measurements available in an image (as opposed to a monopixel detector) each frame contains significant information about the status of the interferometer. Slight pointing errors of the array (equivalent to a phase error between the sub-pupils) can be diagnosed through analysis of the residual starlight distribution.
The frames can be used in a low frequency feedback loop to correct for the non-common path phase delays between the fringe trackers and the detector array: the fringe tracker would introduce a constant phase delay between the beams to correct for these delays.

To remove the residual star light for detection of exoplanets, subtraction of the rotation-invariant component of the light distribution on the detector is a robust approach. Thanks to the high information content of each frame, small variations of the central star image can be modelled in order to further improve this technique. Stellar disk elongation (for rapidly rotating stars seen ``edge-on'') and variations of the null are the two major potential causes of such variations.

\subsection{Final image reconstruction}
A final image can be created from the snapshots using rotational aperture synthesis (\cite{guy01}). However, the presence of a coronagraphic device degrades the quality of the reconstruction: the Fourier transform of a frame does does not yield ``pure'' Fourier components of the real object image. The frame is affected by the coronagraph and no simple exact relation exists between the frame and the object.
This ``degradation'' of the reconstructed image becomes more serious as the number of telescopes is reduced. We believe it becomes acceptable for an array of 6 telescopes or more: for a 6 telescopes array, this effect decreases the S/N by a factor of 3.5 (\S 5.3). Although this factor is quite large, for arrays of 6 or more telescopes, the efficient use of the photons and good imaging capabilities makes this concept very attractive when compared to ``monopixel nullers''. Since the formation of the image is a linear process with a field-dependent PSF it should be possible to correct for this effect. However, the construction of a deconvolution algorithm specific to this problem has not been explored in this work. Figure 8 shows a reconstructed image illustrating this effect.

\section{Planet detection capabilities}

\subsection{Image reconstruction for complex planetary systems}
In this section, we illustrate the potential of our concept at recovering images of complex planetary systems by using a non-redundant array of 6 apertures optimized for (u,v) plane coverage. Rotation of the array is used to fully cover the (u,v) plane.
We simulated the observation of a solar system at a distance of $10\:pc$. Zodiacal light and exozodiacal light have not been included in this simulation. Photon noise was not taken into account (infinite exposure time). The characteristic values for the planets in this model are given in Table 3.
\begin{table}[h]
{\small
\begin{tabular}{|c|c|c|c|c|}
\hline
 & $\theta$(mas) & Teff(K) & $F_{Jy}$ & $F_{ph}$\\
\hline
Sun & 0 & & 3.2 & 5 $10^{-6}$\\
\hline
Mercury & 39 & 480 & 5.8 $10^{-7}$ & 0.91\\
\hline
Venus & 72 & 250 & 2.2 $10^{-7}$ & 0.35\\
\hline
Earth & 100 & 300 & 6.4 $10^{-7}$ & 1.0\\
\hline
Mars & 152 & 270 & 1.0 $10^{-7}$ & 0.16\\
\hline
Jupiter & 520 & 140 & 3.3 $10^{-7}$ & 0.52\\
\hline
Saturn & 950 & 110 & 1.4 $10^{-8}$ & 0.022\\
\hline
Uranus & 1914 & 68 & 7.7 $10^{-13}$ & 1.2 $10^{-6}$\\
\hline
Neptune & 3000 & 55 & 5.0 $10^{-15}$ & 7.7 $10^{-9}$\\
\hline
\end{tabular}
}
\caption{
Values adopted in the simulation for the angular separation $\theta$ to the sun, effective surface temperature (at $10 \mu m$) and flux ($F_{Jy}$ in Jy and $F_{ph}$ in $ph.s^{-1}.m^{-2}.\mu m^{-1}$) for a solar system as seen from $10\:pc$ away. \label{tbl-3}}
\end{table}

\begin{figure}
\centering
\includegraphics[width=9cm]{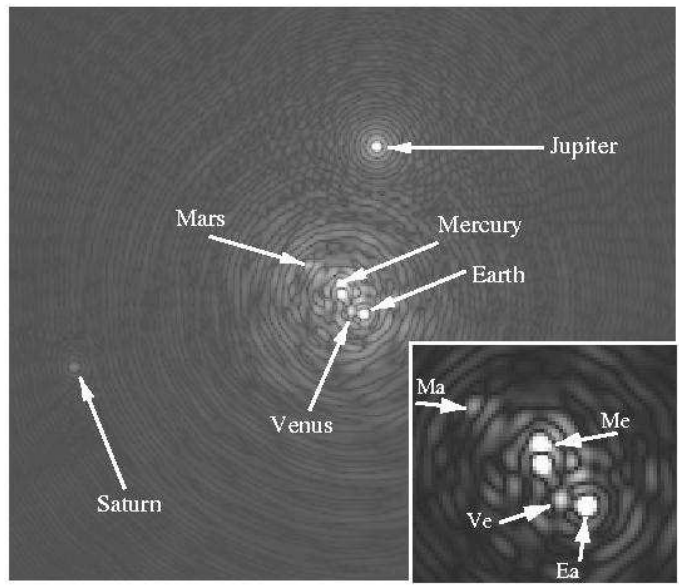}
\caption{Simulated image of the solar system's planets at a distance of $10\:pc$. The values adopted for angular separations and fluxes are given in Table 3. No photon noise has been included. The image of the Sun was numerically substracted.}
\end{figure} 

\begin{figure}
\centering
\includegraphics[width=9cm]{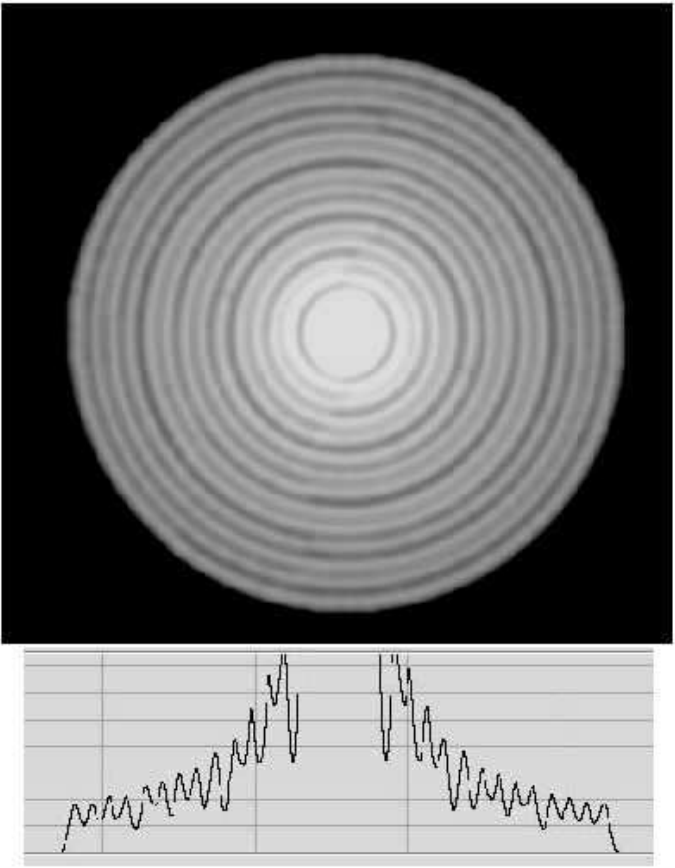}
\caption{Map of the (u,v) coverage obtained after half a rotation of the array. A cut along the diameter of this disk is also shown (linear scale).}
\end{figure} 

Figure 9 shows the (u,v) plane coverage of the observation (half a rotation of the array). The reconstructed image, in which the image of the sun has been subtracted, (Fig. 8) demonstrates the ability of this concept to image planets close to the null (Mercury) as well as planets further out (Saturn). The good quality of the image reconstruction over a wide range of distances allows unambiguous detection of the 6 brightest planets (Uranus and Neptune are very faint due to low effective temperature at $10 \mu m$). Arcs of exozodiacal light and other resolved structures could also be imaged accurately.
The residual ``cloud'' of speckles is due to the the effect of the coronagraph, which corrupts the (u,v) plane values obtained on each snapshot. We have not attempted to remove those predictable artifacts from the image by deconvolution. Because we simulated a system with no exozodiacal light, the effect of this non-perfect reconstruction algorithm is easily visible, since, without the use of a coronagraph (and without a central star), the background between the planets should be null.

\subsection{Theoritical detection sensitivity in imaging mode}
In Appendix B, we give an theoritical estimate of the point source detection sensitivity in a snapshot frame. In this paragraph, we use this estimate to quantify the point source detection sensitivity in imaging mode.
The estimate in Appendix B was computed for a rotation angle of the interferometer such that the planet's image is not occulted by the coronagraph. In a real imaging observation (series of snapshots as the interferometer rotates), the planet's image is periodically occulted by the coronagraph. While on some snapshots the S/N of the detection is given by equation B.23, on others, the planet's signal is partially rejected by the coronagraph and the S/N of the detection is lower. Figure 10 shows the average transmission, $T$, of the interferometer as a function of distance to the optical axis, $x$. At a distance of $0.1''$, the average transmission is 50\%.
\begin{figure}
\centering
\includegraphics[width=6cm,angle=-90]{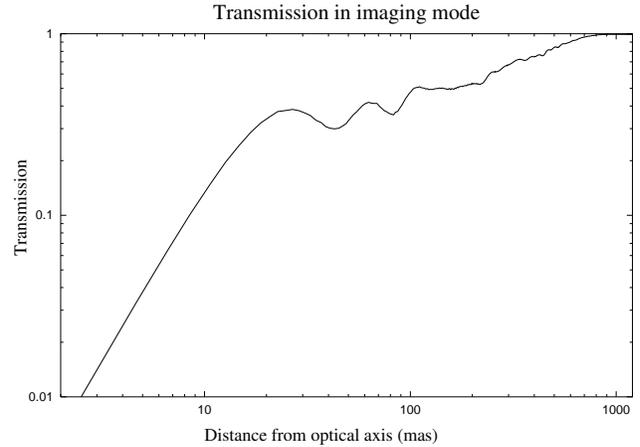}
\caption{Transmission of the interferometer as a function of distance to the optical axis in the imaging mode.}
\end{figure} 
On the final reconstructed image, the planet's flux is the sum of the planet's flux in each snapshot. Following the notations adopted in Appendix B, since this average transmission factor affects $N_{pl}$, $N_{Z}$ and $N_{EZ}$, but not $N_{st1}$ and $N_{st2}$, the expression for the signal to noise of a point source detection becomes
\begin{equation}
\frac{S}{N}(x) = \sqrt{T_{eff}} \frac{N_{pl} \times T(x)}{\sqrt{ (N_{pl}+N_{Z}+N_{EZ})T(x)+N_{st1}+N_{st2} }}. 
\end{equation}
With $T(x)=0.5$ and the values computed in Appendix B, the exposure time required to reach a given S/N is 3.67 times longer than the exposure time required to reach the same S/N in a snapshot in which the planet is not occulted by the coronagraph. Therefore, in imaging mode, with the non-redundant six $3\:m$ apertures, detection ($S/N = 3$) of the Earth at $10\:pc$ is achieved in $29 \: mn.\mu m$ ($5 \: h.\mu m$ with an 0.1 efficiency). In this work, exposure times will be given in $time \times bandwidth$ units, such as $s.\mu m$, corresponding to a $1 \mu m$ bandwidth. $1 s.\mu m$ corresponds to a $1 s$ exposure time for a $1 \mu m$ bandwidth, or a $0.2 s$ exposure time for a $2 \mu m$ bandwidth.

$T(x)$ is higher, except in the central null, when the number of telescopes is larger.

\subsection{Numerical simulation with photon noise}
For a solar system at $10\:pc$, we have identified in Appendix B that the major source of noise is the photon noise from the main star residual light (due to partial resolving of its disk). The effect of ``speckle'' noise, arising from variations in the wavefront of individual beams and OPD variations (this definition includes pointing errors) will be kept small :
\begin{itemize}
\item With a wavefront accuracy of $25 nm$, the dominant source of light contamination arises from the partial resolving of the star's disk (Appendix B). Wavefront stability better than $25 nm$ result in speckles significantly weaker than the stellar light's residual.  
\item Thanks to the rotation of the array, the ``speckles'' will be decorrelated from the planet's signal (this is not true in classical imaging with telescopes). This effect is described in \S 4.2.
\item Postprocessing of individual frames allows, to some extend, to identify wavefront errors (\S 4.2).
\item We propose to use telescopes of modest size (3-meter diameter), which will reduce the number of actuators (and their stroke) needed to actively control the wavefronts. This will keep the amplitude of wavefront variations small.
\end{itemize}

This photon noise is left after subtraction of the central star light on each snapshot and alters the estimation of the (u,v) components of the image and creates noise in the final reconstructed image. This noise appears as speckles with a maximum spatial frequency equal to the maximum spatial frequency of the signal. Figure 11 illustrates this effect: this simulated image of the solar system at $10\:pc$ is to be compared with Fig. 8. In this example, the simulator was used to generate a sequence of 100 noisy snapshots of the solar system and a sequence of 100 noisy snapshots of the sun. Between each snapshot, the array was rotated by $\frac{1}{200}$ of a full rotation, resulting in a rotation of half a turn between the first and the last snapshot of the sequence. Each snapshot is a $972 s \mu m$ exposure, so that the total integration time is $27 h \mu m$. Each of the two images (Solar system and Sun) was reconstructed using the technique presented in \S2.4. The techniques that can be used to reconstruct an image of the star without the planets, using the same set of snapshots, are presented in section \S4.2. The difference between the two images is shown on Fig. 11.

The Earth is the brightest point in this image, and the S/N of the Earth detection, as measured on this reconstructed image, is 10. Jupiter is easily visible, Mercury and Venus (S/N = 3) are seen at the detection limit, but Mars is lost in the noise ($S/N < 2$). In this particular example, the exposure time needed to reach $S/N=10$ for the detection of the Earth in snapshot mode is, from equation B.23, $812 \: s.\mu m$. The non-redundancy of the array is responsible for a multiplication by 4.7 of this exposure time, while the absorption of the Earth's light by the coronagraph in some of the snapshots is responsible for another multiplication by 3.67. In this simulation, photon noise was also simulated to generate the reference image of the Sun without planets. This multiplies the noise in the final subtracted image by $\sqrt{2}$, and therefore requires an exposure time twice as long to reach the same S/N. When all these factors are taken into account, the expected exposure time to reach $S/N = 10$ is $7h47mn$ for a $1 \mu m$ bandwidth (100\% efficiency).

As can be seen in the image, most of the flux of the Earth is not in the narrow diffraction peak, but is diffused in a wide ``cloud'' : this effect is due to the coronagraph and the use of a reconstruction algorithm that does not take it into account (lack of deconvolution). This explains why the exposure time required to reach $S/N=10$ is 3.5 times longer than predicted above. Because this effect is predictable, a proper reconstruction alogrithm (deconvolution) could recover part of this loss. In this example, the combined side-effects of the coronagraph (lower transmission of the planet's photons and the use of an image reconstruction algorithm which does not take it into account)  are responsible for an increase of the exposure time by a factor 12.8. Although this factor is high, it is a small loss compared to the gain that coronagraphy brings: without the coronagraph, the exposure time would have to be $190 days$ ($1 \mu m$ bandwidth) to reach the same S/N. Moreover, some of this loss can be recovered by using a reconstruction algorithm that takes into account the coronagraph, and this loss is decreasing rapidly with the number of apertures.
 
\begin{figure}
\centering
\includegraphics[width=9cm]{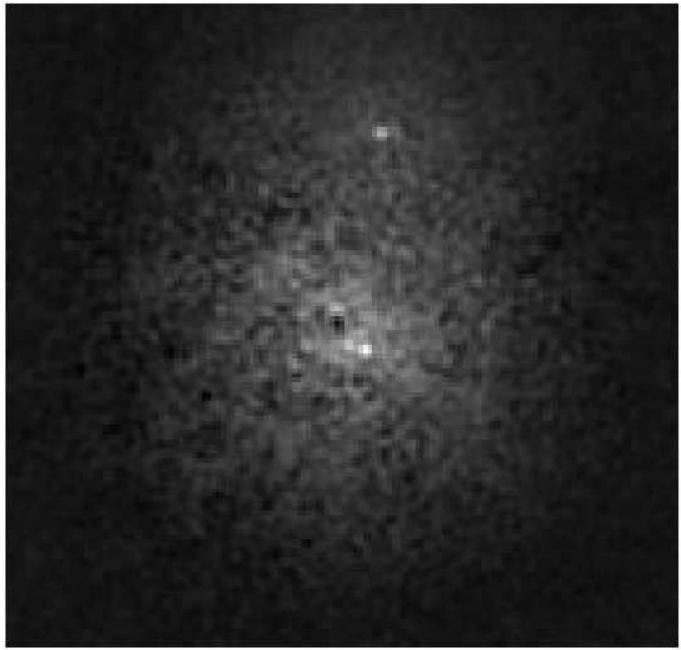}
\caption{Simulated reconstructed image of the solar system at $10\:pc$. Photon noise included. Total exposure time ($T_{eff}$) is $27 h$ (with a $1 \mu m$ bandwidth).}
\end{figure}

\subsection{Spectroscopy}
In this concept, the light of the companion is spread in many diffraction peaks on the focal plane detector for each snapshot exposure. To obtain a spectrum of the companion, we have identified 2 possible strategies:

{\bf Use of a wavelength-sensitive detector array}\\
If each pixel of the detector array used in the focal plane of our concept is sensitive to the wavelength of the incoming photons, it is possible to process the series of frames independently in wavelength bins. Spectral information is then accessible for each source in the field. Arrays of Superconducting Tunnel Junctions (STJs) seem to be the ideal detector for this purpose. STJs detectors are photon-counting devices, and are used to measure the wavelengths of individual photons. An experiments (\cite{pea98}) has demonstrated their ability to measure the wavelengths of photons from the UV to the NIR, with a typical spectral resolution of 10 (spectral resolution decreases with wavelength). A six by six pixels array of STJs has already been used for astronomical observations (\cite{ran00,deb02}). Development of larger arrays at longer wavelengths with a spectral resolution of about 30 at $10 \mu m$ would offer our concept the ideal focal detector for spectroscopic imaging. It is however presently unclear if this technology will be used at such long wavelength, because of the requirement for a very low energy gap supraconductor.

{\bf Dispersing the companion's light on a detector array}\\
Spectroscopy of a point source in the field could be achieved by spatial selection of the source's light in the focal plane. If the position of the source in the sky is previously known, the position of the diffraction peaks of its image in the focal plane can be computed. A mask can then select the source's photons and they would then be spectrally dispersed on a 1D array. The mask would be a regular 2D set of holes if the array is redundant.

A possibly easier way to achieve the same result would be to produce an image without redilution of the pupil to concentrate the source's photons in one single diffraction peak. The light in this diffraction peak would then be spectrally dispersed on a 1D detector. In either case, it would be more efficient to simultaneously acquire on-source and off-source spectra to correct for variation of the null and of the background. If the array' sub-apertures are along a single axis (in the entrance pupil and exit pupil), the light in the focal plane can be concentrated in a narrow ``slit'' (by using cylindrical optics) and then dispersed on a 2D detector array.

Because the exposure time required to acquire a spectrum is significantly longer that the exposure time required to image the planet, we assume that a spectrum is being recorded for a previously imaged planet. Since the position of the planet is known, it is preferable to keep the array in a rotation angle chosen so that the planet is not significantly affected by the coronagraph. We can thereferore use the result of Appendix B to compute the sensitivity of the concept in spectroscopic mode.
With a 0.1 overall efficiency (including the loss of efficiency due to the phase mask, discussed in \S5.2), a spectrum (S/N = 3) with a resolution of 30 can be obtained in 45 minutes for a redundant array and about 3.5 hours for a non-redundant array. With the redundant array, a spectrum with a resolution of 30 and a S/N of 30 can be obtained in 3 days.

\subsection{Discussion}
The S/N ratio in this concept is dominated by the effect of the residual star light $N_{st1}$ and the zodiacal cloud component $N_z$ in the example studied. As can be seen on fig. 12, the relative contributions of those two terms change with $l$, the distance to the system. When the system is close-by, the main source of noise is the residual star light which leaks through the coronagraph because of the large angular diameter of the stellar disk. This leakage decreases rapidly with $l$ ($N_{st1} \propto l^{-4}$) while $N_z$ is independent of $l$. For $l>\:14pc$, the zodiacal component dominates the noise. $N_z$ is only a function of the local zodiacal light background and the wavelength (Equation B.13), while $N_{pl}$, $N_{st1}$ and $N_{st2}$ increase linearly with the total collecting surface of the interferometer (all other things kept equal). For $l>14\:pc$, the sensitivity of the concept is driven by the ratio $N_{pl}/N_{z}$, which can only be increased by increasing the total collecting surface of the array (increasing $N$ or $d$). For $l<14\:pc$, decreasing the baseline $B$ would improve the detection performance, but at the expense of reducing the maximum $l$ for which a system can be detected (about $30\:pc$ with $B=60\:m$).
It is interesting to note that for $l>25\:pc$, $N_{st2}>N_{st1}$: the angular diameter of the star is becoming small enough for the wavefront errors ($\lambda/50$ at $0.5\:\mu m$ in this example) to be the largest source of leakage. If the total collecting area is increased, then this could become the largest source of noise for $l>25\:pc$.
The exozodiacal light component is much lower than either $N_{st1}$ or $N_z$ at all values of $l$. In this example, it would significantly lower the S/N only if the exozodiacal component is more than 20 Zodis. At such concentrations of exozodiacal dust, the spatial structure (clumps, arcs etc...) of the cloud would be a serious issue.
This important conclusion is valid regardless of the type of interferometer/null : this is a general law that applies to all concepts of total collecting surface $42\:m^2$ : $N_z$ dominates $N_{ez}$ for cloud contents less than about 20 Zodis.

\begin{figure}
\centering
\includegraphics[width=6cm,angle=-90]{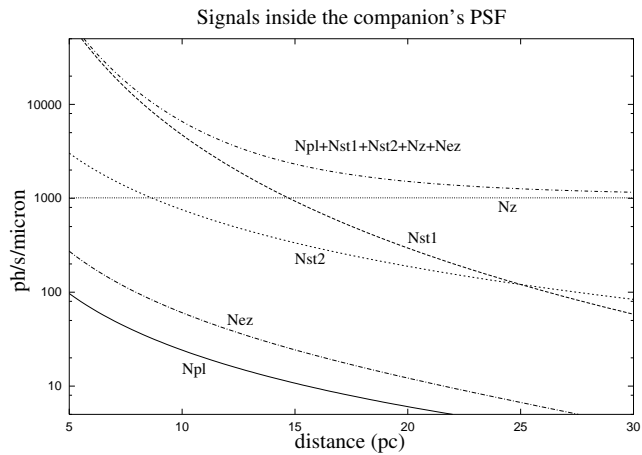}
\caption{Contribution (in $ph.s^{-1}$) of the companion (Npl), the star leakage (Nst1, Nst2), the zodiacal light (Nz) and the exozodiacal light (Nez) to the photon count inside the companion's diffraction peaks, as a function of the distance of the observed planetary system.}
\end{figure} 

\subsection{Comparison with other proposed techniques}
Other proposed techniques use monopixel detectors combined with time-dependent optical pathlength differences. The flux received by the pixel is the product of the brightness map of the object (Star+planet(s)+zodiacal light+exozodiacal light) by a ``transmission map'' determined by the geometry of the interferometer. The transmission map is modified (rotation for example) so that the light of the companion can be extracted. Chopping can be used to improve the sensitivity by increasing the frequency of the planet's signal on the pixel. This synchronous monopixel detection requires a stable and deep null. It is also sensitive to spatial anisotropy of the exozodiacal cloud, because of the poorer imaging capability.
In our imaging concept, on each snapshot, the ``on-source'' and ``off-source'' channels are simultaneously measured (on different pixels of the detector). It is therefore easier to correct for varying background level and varying null efficiency. For example, a variation in nulling performance from frame to frame can be computed and corrected for by analysis of the snapshots: its signature on the detector is different from the planet's PSF.
In our concept, we combine nulling and spatial decoupling of the planet's photons and the star's photons on the detector, whereas other concepts only use one of those 2 techniques (nulling for interferometers and spatial decoupling for single aperture imagers). In this work, we have demonstrated how combining the 2 techniques lowers the technical requirements characteristic of each of the techniques:\\
(1) The central null does not need to be in $\theta^4$ or $\theta^6$. A $\theta^2$ null is sufficient.\\
(2) The tolerance for wavefront errors and cophasing errors are lower: $\lambda/20$ at $0.5\:\mu m$ for detection of an Earth at $10\:pc$.\\
In most nulling interferometer concepts, the null relies on a particular geometry of the array.
In this concept, however, the null can be achieved with any array configuration, thus offering the possibility of a very good (u,v) plane coverage and excellent wide field imaging capabilities. Complex exozodiacal structures can be imaged and the risk of confusion between a planet and such structures is much lower.
Imaging concepts using pupil densification and phase mask coronagraphy have been proposed by Boccaletti et al. (\cite{boc00}) and Guyon \& Roddier (\cite{guy00}). In the concept presented by Boccaletti et al., there is no redilution of the pupil, and the authors suggest the use of a high number of apertures ($N=36$) to obtain a clean field of view large enough for exoplanet detection. In the Guyon \& Roddier concept, a small number of apertures ($N=6$) is used along with pupil redilution for wide field of view imaging. In both concepts, the imaging capabilities were very limited, and aperture synthesis was not explored. The S/N computations in Appendix B also apply to those two concepts.

\subsection{Implementation - Complexity}
Unlike most nulling beam-combiners, the number of optical elements in our Fizeau imager does not increase rapidly with the number of sub-apertures. Figure 13 shows a possible optical layout for this concept: in this figure, there are a total of $6 \times N + 3$ reflecting surfaces and one transmissive element (the apodization mask, which could also be made reflective), excluding delay lines and at least one beam splitter (probably in the densified pupil) for wavefront sensing and fringe tracking. M1, the primary mirror, focuses the light on M2 which, with M3, is used to image the telescope pupil on the apodization mask. M3 also collimates the beam. Optics of the delay lines have not been represented in this simple figure. The M4 mirrors pick up the beams to create the densified pupil. The M5 parabolic mirror forms an image on the phase mask and recollimates the beam before pupil redilution. The M6 and M7 mirrors redilute the pupil before M8, the final imaging mirror.
The nulling and beam combination optics are very simple and only have $2 \times N +2$ reflecting surfaces and one transmissive mask. For $N>4$ This number of optical elements is lower than for a non-coronagraphic nuller, for which the number of beam splitters required is about $N^2/2$. This solution seems especially attractive for large numbers of apertures.
\begin{figure}
\centering
\includegraphics[width=9cm]{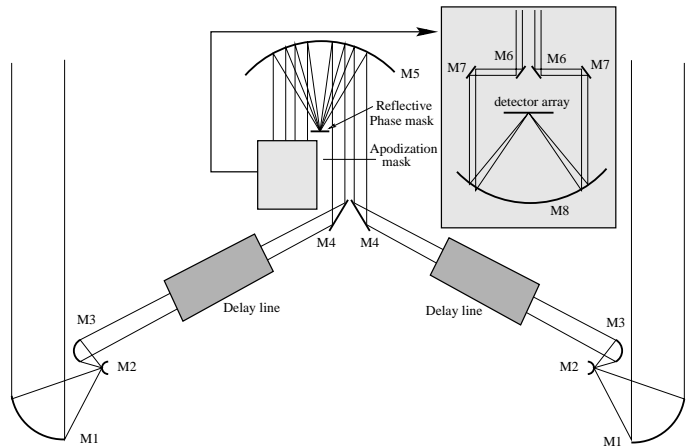}
\caption{Schematic of a possible optical layout for the concept.}
\end{figure}

\section{Other capabilities}

\subsection{Non-coronagraphic imaging at $10 \:\mu m$}
When the phase mask coronagraph is removed from the focal plane between the 2 densified pupils, the interferometer is an aperture synthesis imager. With six 2-meter apertures in a non-redundant configuration, this concept can have a full (u,v) plane coverage (up to 60m baseline) with half a rotation of the array.
The field of view of the reconstructed image is only limited by the optics and detector, and can be several times $\lambda/d$. With $10\:mas$ pixels, a $2048 \times 2048$ detector array would offer a $20\:''$ field of view. The angular resolution at $10\: \mu m$ would be $35\: mas$.
No modification of the optics, other than moving the phase mask out of the focal plane of the densified pupil, would be required.

\subsection{Imaging in the visible or Near-IR}
The wavefront errors acceptable for imaging ($\lambda/10$) are much larger than for nulling ($\approx \lambda/500$). Achieving the wavefront correction required for nulling at $10\: \mu m$ allows the interferometer to do imaging down to visible wavelengths. In this concept, most of the visible photons would probably be used for wavefront sensing and aperture cophasing, and the reflectivity of IR-optimized optics might not be very high in the visible. For those reasons, the sensitivity of the interferometer would probably be much below the theoretical sensitivity of a $42\: m^2$ collecting area (six $3\:m$ diameter apertures), but the angular resolution at $0.5\: \mu m$ would be less than $2\: mas$ ($60\:m$ baseline). A dedicated visible detector array would be needed to take advantage of this potential. With a 10k x 10k array with $1\:mas$ per pixel, the field of view would be $10\:''$.
The scientific interest of non-coronagraphic imaging would be greater for objects significantly fainter, at visible wavelength, than close-by ($d < 20\:pc$) stars. For example, the apparent visible luminosity of bright AGNs and quasars is typically 100 times less that a Sun-like star at $20\:pc$. This reduced number of photons for wavefront sensing and apertures cophasing could increase wavefront/cophasing errors and bring the shortest possible imaging wavelength form the visible to the Near-Infrared.

\subsection{Off-source imaging}
The very wide field of view theoretically achievable by this concept makes it possible to have both a wavefront sensing source and a science target in the same field of view. It is possible to obtain a reasonable sky coverage without relying on dual-beam interferometry, which would considerably increase the complexity of the interferometer. With a limiting visible magnitude of 15 (similar to the limiting magnitude of adaptive optics systems on 3-meter telescopes), and an off-axis guiding range of up to $20\:''$, the sky coverage is 10 \%.
If the science target and the guide star are separated by more than $\lambda/d$ ($1\:''$ at $10\: \mu m$), the ``mixing'' of the object photons and the guide star photons is very low. If it were an issue, the nulling coronagraph could be used to further reduce this effect.

\section{Conclusion}
The phase mask nulling stellar coronagraph can be applied to interferometric arrays with a small number of telescopes to produce clean wide field of view images. Rotational aperture synthesis allows full (u,v) plane coverage up to a maximum baseline of $60\:m$ with only six 2m-apertures.
The use of an imaging coronagraphic technique is a very powerful tool to detect and image Earth-size planets around other stars. Exposure times of a few hours are sufficient to image an Earth-like planet around a Sun-like star. A planet is easily distinguished from an exozodiacal disk thanks to this imaging technique. The sensitivity to background emission (thermal emission of the optics, zodiacal light) and residual star light is very good because of the high spatial selectivity of the image. This high spatial selectivity also allows good spectroscopic sensitivity, which could be fully exploited with the use of STJ detector arrays.
The wide field of view and good imaging capability of this concept also make it a very powerful instrument for general astrophysics.
This solution is very attractive for simplicity of the optical layout especially for large number (more than 10) of apertures.

\begin{acknowledgements}
The authors are very grateful to the Boeing-SVS TPF study group for many helpful discussions that helped improve this concept. We also thank Catherine Ishida and Toby Owen for their suggestions to improve the manuscript. This work was supported by JPL grant 1217288, awarded to Boeing-SVS by JPL to study possible concepts for the TPF mission. 
\end{acknowledgements}

\appendix
\section{Performance of the phase mask coronagraph}
\subsection{The theory of the phase mask coronagraph for a circular aperture}
The phase mask coronagraph (\cite{rod97,guy99}) uses a small mask to dephase the light in the central part of the Airy pattern by half a period. The mask is a disk covering 43\% the diameter of the first dark Airy ring. As shown on Fig. A.1, in the focal plane of the telescope, the light complex amplitude is multiplied by -1 inside the phase mask (phase shift of half a period) while it remains unchanged outside the mask.
\begin{figure}
\centering
\includegraphics[width=9cm]{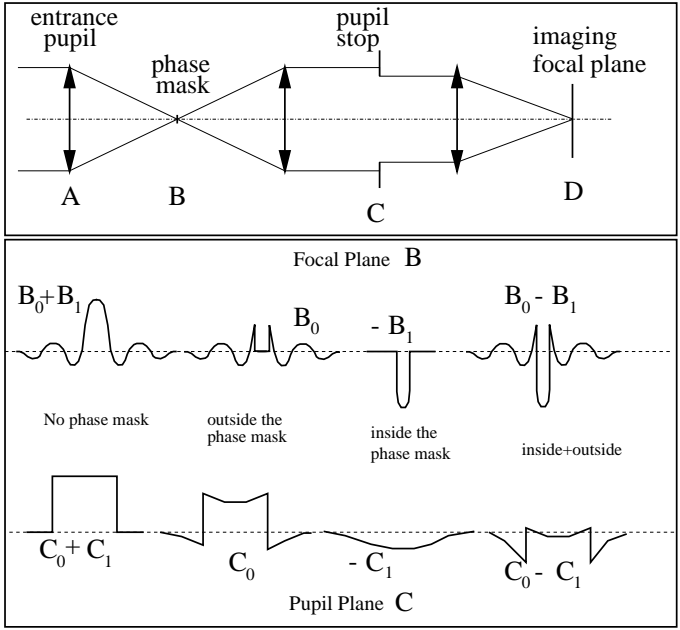}
\caption{The phase mask coronagraph for a circular aperture.}
\end{figure} 

With $\vec{\alpha}$ and $\vec{\beta}$ the position vectors in the focal and pupil planes respectively,  and $A(\vec{\beta})$, $B(\vec{\alpha})$, $C(\vec{\beta})$ and $D(\vec{\alpha})$ the light amplitude functions in planes A, B, C and D of Fig. A.1, 
\begin{equation}
B(\vec{\alpha}) = \mathscr{FT}(A(\vec{\beta})),
\end{equation}
where $\mathscr{FT}$ denotes the Fourier transform (and $\mathscr{FT}^{-1}$ the inverse Fourier Transform). For a circular entrance pupil of radius R, $A(\vec{\beta})=1$ if $|\vec{\beta}|<R$, and $|B(\vec{\alpha})|^2$ is the Airy function. When the phase mask is introduced :
\begin{equation}
B(\vec{\alpha}) = B_0(\vec{\alpha}) - B_1(\vec{\alpha}).
\end{equation}
$B_0(\vec{\alpha})$ and $B_1(\vec{\alpha})$ are the complex amplitudes outside and inside the area covered by the phase mask, respectively. The effect of the phase mask is to multiply $B_1(\vec{\alpha})$ by -1. With the same notation:
\begin{equation}
C(\vec{\beta}) = \mathscr{FT}^{-1}(B(\vec{\alpha})) = C_0(\vec{\beta})-C_1(\vec{\beta}),
\end{equation}
\begin{equation}
C_0(\vec{\beta}) =\mathscr{FT}^{-1}(B_0(\vec{\alpha})),
\end{equation}
\begin{equation}
C_1(\vec{\beta}) =\mathscr{FT}^{-1}(B_1(\vec{\alpha})).
\end{equation}
Increasing the phase mask radius increases $C_1(\vec{\beta})$ and decreases $C_0(\vec{\beta})$ inside the pupil. For a critical value of the phase mask radius, $C(\vec{\beta})$ is then brought close to zero inside the pupil, as shown on Fig. A.1. A ``Lyot stop'' that blocks the light outside the pupil in plane C is used to reject the light diffracted outside the pupil. For an off-axis point source, the phase mask in focal plane B has no effect (the Airy pattern is far from the mask) and all the light is inside the pupil in plane C: there is very little extinction for an off-axis source. With this simple phase mask coronagraph design, the maximum extinction factor (monochromatic on-axis point source, optimal phase mask size) for a circular pupil is 160. This simple design has been successfully tested in an experiment we carried out on an optical bench with a monochromatic light source (Guyon et. al., 1999). 

\subsection{Entrance pupil apodization}
When the phase mask size is optimal, the extinction factor for an on-axis point source cannot be better than 160 for a uniformly lit circular pupil. This can be seen in Fig. A.1: $C(\vec{\beta})$ is not flat inside the pupil but has a positive curvature on this representation. If the entrance pupil, $A(\vec{\beta})$ has a negative curvature (transmission increases towards the center of the pupil), the positive curvature of $C(\vec{\beta})$ can be canceled. By a simultaneous optimization of the transmission map of the entrance pupil and the size of the phase mask, it is possible to reach a total extinction for an on-axis point source.

We have developed an algorithm that computes the optimal transmission map for the pupil by changing $A(\vec{\beta})$ according to the $C(\vec{\beta})$ residual inside the pupil: three iterations bring the theoretical extinction factor from 160 to $10^7$ (monochromatic light, on-axis point source). For a circular entrance pupil, the transmission map has a 49\% minimal transmission at the edges of the pupil and a total integrated transmission of 73\%. The apodization mask for such a pupil can be seen on Fig. A.2 (left). The phase mask size has to be increased when the apodization mask is used. Such apodization masks have been studied in great detail in the case of rectangular pupils (\cite{aim01,aim02}). In the following discussion, we will consider an apodized entrance pupil.

\begin{figure}
\centering
\includegraphics[width=9cm]{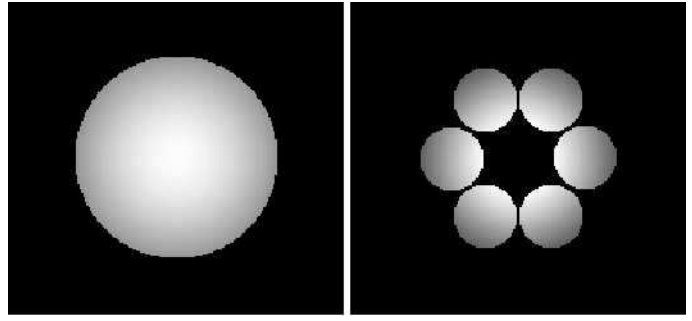}
\caption{Apodized entrance pupils for the phase mask coronagraph : single circular pupil (left) and interferometer's densified pupil (right).} 
\end{figure} 

The algorithm used to compute the apodization mask for a single aperture telescope can also be used in a multi-aperture pupil. The result of this algorithm for our 6 telescope interferometer densified pupil is shown on Fig. A.2 (right). It is interesting to note that even for an unfilled pupil, this apodization mask yields a infinite theoretical extinction for an on-axis point source. However, as the filling factor of the entrance pupil goes down, the total integrated transmission of the apodization mask decreases. For our 6 telescope interferometer densified pupil, the minimum transmission is 23\% on the outside edges of the densified pupil and the total integrated transmission is 57.3\%. Had the sub-pupils been hexagonal, the total transmission would have been 66\% thanks to a better filling factor. In this study, we will consider an ideal nulling coronagraph in which the chromatism problem has been solved, equipped with an apodization mask in the entrance pupil plane.

In this appendix, we also estimate the amount of residual light in the coronagraphic focal plane arising from different errors.
\subsection{The phase mask thickness}
The relation between the phase shift $p$ and the thickness $e$ of the mask is given by

\begin{equation}
\label{equ1}
p = (n-1) \frac{e\:2\pi}{\lambda} 
\end{equation}

$n$ being the index of the material used to make the mask.
A phase mask thickness error of $\delta e$ yields a phase error
\begin{equation}
\delta p = (n-1) \frac{\delta e\:2\pi}{\lambda}
\end{equation}
For an on-axis point source, and an apodized entrance pupil, the residual light amplitude distribution in the pupil plane is the Fourier transform of the complex amplitude error in the focal plane.
\begin{equation}
Res(\vec{\beta}) = \mathscr{FT}[(n-1) \delta e \times B_1(\vec{\alpha})]
\end{equation}
The corresponding light contribution in the focal plane is
\begin{equation}
L = \int_{|\vec{\beta}|<R}{Res(\vec{\beta})^2}{d}{\vec{\beta}} = \pi^2 (\frac{\delta e}{e})^2 \times \int_{|\vec{\beta}|<R}{(\mathscr{FT}[B_1(\vec{\alpha})])^2}{d}{\vec{\beta}}
\end{equation}
The fraction of the total incoming light of the point source that ``leaks'' into the final focal plane is
\begin{equation}
F_1 = \pi^2 (\frac{\delta e}{e})^2 \times \frac{\int_{|\vec{\alpha}|<R}{(\mathscr{FT}[B_1(\vec{\alpha})])^2}{d}{\vec{\alpha}}}{\int{(\mathscr{FT}[B_0(\vec{\alpha}) + B_1(\vec{\alpha})])^2}{d}{\vec{\alpha}}}
\end{equation}
In the apodized pupil scheme, $\mathscr{FT}[B_0(\vec{\alpha})] - \mathscr{FT}[B_1(\vec{\alpha})] = 0$ inside inside the pupil,
\begin{equation}
F_1 = \frac{\pi^2}{4} \times (\frac{\delta e}{e})^2
\end{equation}
With $(\frac{\delta e}{e}) = 0.001$, $F_1 = 2.5\: 10^{-6}$.

\subsection{The phase mask diameter}
When the mask diameter is changed from $\rho_0$ to $\rho_0+\delta \rho$, the same approach leads to
\begin{equation}
Res(\vec{\beta}) = \mathscr{FT}[2 B(\vec{\alpha})_{\rho_0<|\vec{\alpha}|<\rho_0+\delta \rho}]
\end{equation}
And the fraction of the total incoming light that leaks into the the final focal plane image is
\begin{equation}
F_2 = \frac{\int_{|\vec{\alpha}|<R}{(\mathscr{FT}[2 B(\vec{\alpha})_{\rho_0<|\vec{\alpha}|<\rho_0+\delta \rho}])^2}{d}{\vec{\alpha}}}{\int{(\mathscr{FT}[B_0(\vec{\alpha}) + B_1(\vec{\alpha})])^2}{d}{\vec{\alpha}}}
\end{equation}
for small values of $\delta \rho$
\begin{equation}
F_2 = \frac{\rho_0^2 \times \int_{|\vec{\alpha}|<R}{(\mathscr{FT}[2 B(\vec{\alpha}) \delta(|\vec{\alpha}|-\rho_0)])^2}{d}{\vec{\alpha}}}{\int{(\mathscr{FT}[B_0(\vec{\alpha}) + B_1(\vec{\alpha})])^2}{d}{\vec{\alpha}}} \times (\frac{\delta \rho}{\rho_0})^2 
\end{equation}
where $\delta(|\vec{\alpha}|-\rho_0)$ is one if $|\vec{\alpha}|=\rho_0$ and zero elsewhere. Numerically
\begin{equation}
F_2 = 1.702 \times (\frac{\delta \rho}{\rho_0})^2
\end{equation}
With $(\frac{\delta \rho}{\rho_0}) = 0.001$, $F_2 = 1.7\: 10^{-6}$.

\subsection{The spectral bandwidth}
The phase mask needs to be achromatic : the phase shift should be half a period over the range of wavelength used for the imaging. Such an achromatic phase mask could be realized by stacking several layers of different materials, whose index-wavelength curves have to be carefully chosen. Let's consider an observation integrating wavelengths from $\lambda_0-\delta \lambda$ to $\lambda_0+\delta \lambda$ (flat spectrum) with a phase mask coronagraph optimized for $\lambda_0$. Without an achromatic phase mask, the phase shift introduced by the mask is a linear function of the wavelength, and, by integration of equation A.11. over the bandwidth, we obtain
\begin{equation}
F_3 = \frac{1}{3} \: (\frac{\delta \lambda}{\lambda_{0}})^2
\end{equation}
where $F_3$ is the light ``leakage'' arising from the wavelength dependance of the phase shift.

Another source of chromatism is the variation of the Airy pattern size with wavelength. The change of wavelength is equivalent to a change of the phase mask size and the 2 are linked by 
\begin{equation}
\frac{\delta \lambda}{\lambda_0} = \frac{\delta \rho}{\rho_0}
\end{equation}
From integration of equation A.14, the light ``leakage'' is then 
\begin{equation}
F_4 = \frac{\rho_0^2 \times \int_{|\vec{\alpha}|<R}{(\mathscr{FT}[2 B(\vec{\alpha}) \delta(|\vec{\alpha}|-\rho_0)])^2}{d}{\vec{\alpha}}}{\int{(\mathscr{FT}[B_0(\vec{\alpha}) + B_1(\vec{\alpha})])^2}{d}{\vec{\alpha}}} \times \frac{1}{3} (\frac{\delta \lambda}{\lambda_0})^2
\end{equation}
Numerically,
\begin{equation}
F_4 = 0.567 \times (\frac{\delta \lambda}{\lambda_0})^2
\end{equation}
For a $1\: \mu m$ bandwidth centered on $10\: \mu m$, 0.15 percent of the light of a point source is ``leaking'' into the final image. This problem can be solved at the expense of adding a wavelength-dependent magnification assembly. Such a device, made out of a few chromatic lenses, has already been successfully used to record fringes in white light (Roddier et. al., 1980) and for speckle imaging (Boccaletti et. al. 1998). This is will extend the bandwidth by a factor of at least 100, bringing down the light leak to less than $10^{-5}$ for a $1\: \mu m $ bandwidth at $10\: \mu m$.
Another very interesting solution to solve the chomatism problem, as suggested by Antoine Labeyrie (private communication), is to use a Bragg hologram as a phase mask. The mask could then be made so that each wavelength sees it as being the right size and having the right phase shift.

\subsection{Pointing errors and wavefront errors}
For a pointing error of $\delta \vec{\alpha}$, the light ``leakage'' is, for small values of $\delta \vec{\alpha}$
\begin{equation}
F_5 = \frac{\int_{|\vec{\alpha}|<R}{(\mathscr{FT}[\int{2 B(\vec{\alpha}) \delta(|\vec{\alpha}|-\rho_0) cos(\theta-\theta_{\delta \vec{\alpha}})}{d}{\theta}])^2}{d}{\vec{\alpha}}}{\int{(\mathscr{FT}[B_0(\vec{\alpha}) + B_1(\vec{\alpha})])^2}{d}{\vec{\alpha}}} \times |\delta \vec{\alpha}|^2
\end{equation}
Numerical simulations give, for small values of $\delta \vec{\alpha}$,
\begin{equation}
F_5 = 0.364 \times \frac{|\delta \vec{\alpha}|^2}{\rho_0^2}
\end{equation}
For higher-order phase errors, the computer simulations give
\begin{equation}
F_5 = \gamma \times (1-Strehl)
\end{equation}
where $Strehl$ is the Strehl ratio of the PSF. This formula was tested for Zernikes polynomials 4 to 100 and $\gamma$ is then between $0.6$ and $1$, depending upon the Zernike polynomial.

\section{Point source detection sensitivity in a snapshot exposure}
In this appendix, we compute the point source detection sensitivity without making any assumptions about the image reconstruction algorithm. We consider a snapshot frame in which the distribution of the residual starlight is exactly known. The photons of the point source (planet) are concentrated in a series of diffraction peaks. We estimate the signal (planet's photons) and noise (photon noise from the residual starlight, background, zodiacal light and exozodiacal light) in these diffraction peaks. This computation does not take into account the problem of confusion with other structures (exozodiacal light concentrations, other planets), a problem for which this concept is well suited.
  
The computations are first done for a redundant array and we then give quantitative measures for the decrease of S/N due to the non-redundancy of the array. We suppose that the planet's position on the sky is not on a null of the coronagraph. In a real observation of a peviously unknown planet, there is a probability for the planet to be occulted by the coronagraph. This effect is quantified in \S5.2.

We consider a space interferometer of $N$ telescopes of diameter $d$ spread along a baseline $B$ (end to end), operating at $\lambda = 10 \:\mu m$ (spectral bandwidth of $1\: \mu m$). The observation target is a star of angular diameter $R_{st}$ of light flux $F_{st}$ (in $ph.s^{-1}.m^2.\mu m^{-1}$) as seen from Earth and a companion with a light flux $F_{pl}$ at an angular distance $a$ from the star. At $10 \mu m$, the ratio between $F_{st}$ and $F_{pl}$ for a Sun-Earth system is approximately $5 \times 10^6$.
We give analytical expressions for each of the terms of the S/N expression. We also give numerical values for the observation of a Sun-Earth system at $10\:pc$ ($F_{st}=5 \times 10^6\:ph.s^{-1}.m^{-2}.\mu m^{-1}$,$F_{pl}=1\:ph.s^{-1}.m^{-2}.\mu m^{-1}$,$R_{st}=2.26 \times 10^{-9}\:rad$), with a 1 Zodi cloud (face-on). Whenever numerical values are given, we consider a 6 aperture interferometer, $3\: m$ diameter each with a baseline of $60\:m$ ($N=6$, $B=60\:m$, $d=3\:m$).



{\bf The companion's photon count}\\
The size of a diffraction peak in the PSF is $\lambda/B$. There are $(B/(d \times \sqrt{N}))^2$ such diffraction peaks in the PSF (redundant array).
With $\tau$ the average transmission of the apodization mask in the pupil ($0.57$ for a 6 aperture interferometer), $\tau (F_{pl} \pi d^2)/4$ is the total number of photons from each aperture in the diffraction peak of an aperture. Since the area of this diffraction peak is $(\lambda/d)^2$, $\tau (F_{pl} \pi d^4)/(4 \lambda^2)$ is the peak intensity in the image of the companion by one telescope. Because N telescopes' images are combined coherently, the peak intensity of the recombined image is $N^2$ times the peak intensity of the companion image in one telescope. 
The photon count inside the diffraction peaks of the companion's PSF is obtained by multiplying the photon count per diffraction peak by the number of diffraction peaks:
\begin{equation}
N_{pl} = N^2 \times \tau \frac{F_{pl} \pi d^4}{4 \lambda^2} \times (\frac{\lambda}{B})^2 \times (\frac{B}{d \sqrt{N}})^2,
\end{equation}
\begin{equation}
N_{pl} = N \times \tau \frac{F_{pl} \pi d^2}{4}.
\end{equation}
Because we have made the assumption that the array is redundant, we obtain the expected result that all of the companion's light ``collected'' by the interferometer is in the diffraction peaks of the PSF. The numerical value for our example is
\begin{equation}
N_{pl} = 24.17\: ph.s^{-1}.\mu m^{-1}.
\end{equation}

{\bf The residual light from the star}\\
Because the angular diameter of the Solar-type stars at 5 to $20\:pc$ is 0.5 to $2\:mas$, in this computation, we do not need to worry about the star's angular diameter being larger than the distance between consecutive diffraction peaks in the PSF of the interferometer. The star is also smaller than the angular resolution of the interferometer we consider. Therefore, for a given direction, the residual light from a point source at an angular distance $\theta$ from the optical axis is proportional to $\theta^2$ (Appendix A). When using an array that has no strong preferential direction (unlike the (u,v) plane coverage optimized arrays), the null is symmetric and the off-axis residual flux fraction equation, is, as obtained from our simulations (\cite{guy00}), 
\begin{eqnarray*}
F_R = (1.6 \times \frac{B}{\lambda})^2 \times  \theta^2 \\
for &  \theta < \frac{\lambda}{1.6 \times B},
\end{eqnarray*}
\begin{eqnarray*}
F_R = 1\\
for &  \theta > \frac{\lambda}{1.6 \times B}.
\end{eqnarray*}
In this work, we will consider $R_{st}$ to be smaller than $\frac{\lambda}{1.6 \times B}$.
By integrating this equation over the stellar disk, the residual light from the on-axis star is
\begin{equation}
{F_{st}}_{res} = F_{st} \times (1.6 \times \frac{B}{\lambda})^2 \times \frac{R_{st}^2}{2}.
\end{equation}
The residual light from the star is mostly concentrated in a small ring around the optical axis, and the detection of a companion is thus easier if the angular separation is important. However, for this simple estimate, we will consider that this residual light is spread evenly inside the PSF envelope ($\lambda/d$). This is a very conservative assumption that will yield to pessimistic values for the detection of companions that are not in the bright ring of residual light from the star. The residual star light is spread over $\frac{\lambda}{d}$, as is the light of the companion. The diffraction peaks from the PSF of the companion occupy $\frac{1}{N}$ of this area (redundant array). Therefore, from (B.4), the photon count due to this residual is 
\begin{equation}
N_{st1} = N \tau \frac{\pi d^2}{4} F_{st} (1.6 \times \frac{B}{\lambda})^2 \frac{R_{st}^2}{2} \times \frac{1}{N},
\end{equation}
\begin{equation}
N_{st1} = \tau \frac{\pi d^2}{4} F_{st} (1.6 \times \frac{B}{\lambda})^2 \frac{R_{st}^2}{2}.
\end{equation}
The numerical value for our example is
\begin{equation}
N_{st1} = 4721\: ph.s^{-1}.\mu m^{-1}.
\end{equation}

{\bf The wavefront errors}\\
With the same conservative assumption as for the residual light from the star estimate, and using (A.21), with $\gamma=1$, 
\begin{equation}
N_{st2} = \tau \frac{\pi d^2}{4} F_{st} (1-Strehl).
\end{equation}
In this formula, $Strehl$ is the Strehl of the PSF in the densified pupil scheme, and is to be related to densified pupil wavefront errors, which include cophasing errors between the sub-apertures. For small wavefront errors, the $Strehl$ and the variance of the wavefront phase errors $\sigma^2$ are related through
\begin{equation}
Strehl = e^{-\sigma^2}.
\end{equation}
By inserting equation B.9 into equation B.8, for small wavefront errors,
\begin{equation}
N_{st2} = \tau \frac{\pi d^2}{4} F_{st} \sigma^2.
\end{equation}
$N_{st1} = N_{st2}$ for $\sigma = 1.53 \:10^{-2} rad$, which corresponds to a $\lambda/410$ precision on the wavefront at $10 \mu m$, or $\lambda/20$ at $0.5 \mu m$. Reaching this accuracy on the wavefront is possible with smooth polishing of all optics and good cophasing of the sub-apertures. In this study, we consider a wavefront flat to $\lambda/50$ at $0.5\mu m$, for which $N_{st2} = 0.16 \times N_{st1}$, and we adopt
\begin{equation}
N_{st2} = 755\: ph.s^{-1}.\mu m^{-1}.
\end{equation}

{\bf The instrumental thermal emission and zodiacal light}\\
With $B_{Z}$ the background contribution from zodiacal light ($ph s^{-1} m^{-2} sr^{-1}$), the corresponding photon count ($ph s^{-1}$) inside the companion's PSF diffraction peaks is
\begin{equation}
N_{Z} = N \tau \frac{\pi d^2}{4} \times B_{Z} \times (\frac{\lambda}{B})^2 \times (\frac{B}{d \sqrt{N}})^2 
\end{equation}
\begin{equation}
N_{Z} = \tau \frac{\pi}{4} \times B_{Z} \times {\lambda}^2.  
\end{equation}
We adopt a value of $15 MJy sr^{-1}$ for the zodiacal light component at $10 \mu m$.
At $10 \mu m$,
\begin{equation}
1 MJy sr^{-1} = 1.5 \times 10^{12}\: ph \mu m^{-1} sr^{-1} m^{-2} s^{-1} 
\end{equation}
hence,
\begin{equation}
B_{Z} = 2.25 \times 10^{13} ph \mu m^{-1} sr^{-1} m^{-2} s^{-1},
\end{equation}
and
\begin{equation}
N_{Z} = 1.01\: 10^3 ph.\mu m^{-1}.s^{-1}.
\end{equation}
The thermal emission of the optics of the interferometer (telescopes + recombination optics + detector) can be included in $N_{Z}$, but by cooling the optics below $40\:K$, this contribution is smaller than the zodiacal light.

{\bf Exozodiacal light component}\\
Estimating the amount of exozodiacal light photons in the diffraction peaks of the companion's PSF is harder than what we did for the zodiacal light component because the exozodiacal light cannot be considered as uniform across the field of view of the interferometer. For a given exozodiacal cloud model, the photon count will depend upon the relative position of the companion and the star (center of the exozodiacal cloud).
For this example, we adopt the following model for the exozodiacal cloud (\cite{rea95}):
\begin{itemize}
\item The cloud is seen face-on.
\item The cloud vertical optical depth decreases as $r^{-0.37}$, $r$ being the distance from the star in Astronomical Units.
\item The grain temperature decreases as $r^{-0.42}$.
\item The inner edge of the cloud is at $r=0.02$ (in the solar system, distance at which the dust temperature is 1300 K, the dust sublimation temperature).
\end{itemize}
We adopt a value of 25 MJy/sr at $r=1$ for a 1 Zodi disk, and we consider a $z$ Zodi disk. With this model, the surface brightness of the exozodiacal cloud is, in MJy/sr,
\begin{equation}
EZ(r) = \frac{1117.53 \times z \times r^{-0.37}}{e^{4.8 \times r^{0.42} - 1}},
\end{equation}
or, in $ph.s^{-1}.\mu m^{-1}.m^{-2}.sr^{-1}$,
\begin{equation}
EZ(r) = \frac{1.676\: 10^{15} \times z \times r^{-0.37}}{e^{4.8 \times r^{0.42} - 1}}
\end{equation}
for $r>0.02$, and $0$ for $r<0.02$.
In the final image, the companion light is concentrated in a regular pattern (redundant array) of diffraction peaks, separated by $\frac{\lambda \times \sqrt{N}}{B}$ one from the other in each axis. There are also multiple images of the exozodiacal cloud, each separated by the same distance $\frac{\lambda \times \sqrt{N}}{B}$ on each axis. The shift between these two regular patterns is given by the position of the companion relative to the star. Because of the periodicity of these patterns (redundant array), the number of exozodiacal photons in the companion's PSF is a 2-D periodic function of the relative position of the companion and the star.
By summing the exozodiacal cloud surface brightness over the number of diffraction peaks ($\frac{B^2}{d^2 \times N}$) of the companion's PSF and the number of images of the exozodiacal cloud, we can compute $N_{EZ}$, the number of exozodiacal photons on the companion's PSF diffraction peaks 
\begin{equation}
N_{EZ} = N \tau \frac{\pi d^2}{4}
\times (\frac{\lambda}{B})^2 
\times B_{EZ},
\end{equation}
where $B_{EZ}$ is the Exozodiacal light Equivalent background level (in $ph.\mu m^{-1}.s^{-1}.m^{-2}.sr^{-1}$) and
\begin{equation}
B_{EZ} = \sum_{k,l} EZ(\sqrt{(x+k \times \delta)^2+(y+l \times \delta)^2}),
\end{equation}

where $k$ and $l$ are integers which are used to compute the discrete points of the exozodiacal cloud wich will contribute to $N_{EZ}$ and $\delta$ is the distance between successive diffraction peaks,

\begin{equation}
\delta = \frac{\lambda \times \sqrt{N}}{B}.
\end{equation}

\begin{figure}
\centering
\includegraphics[width=12cm,angle=90]{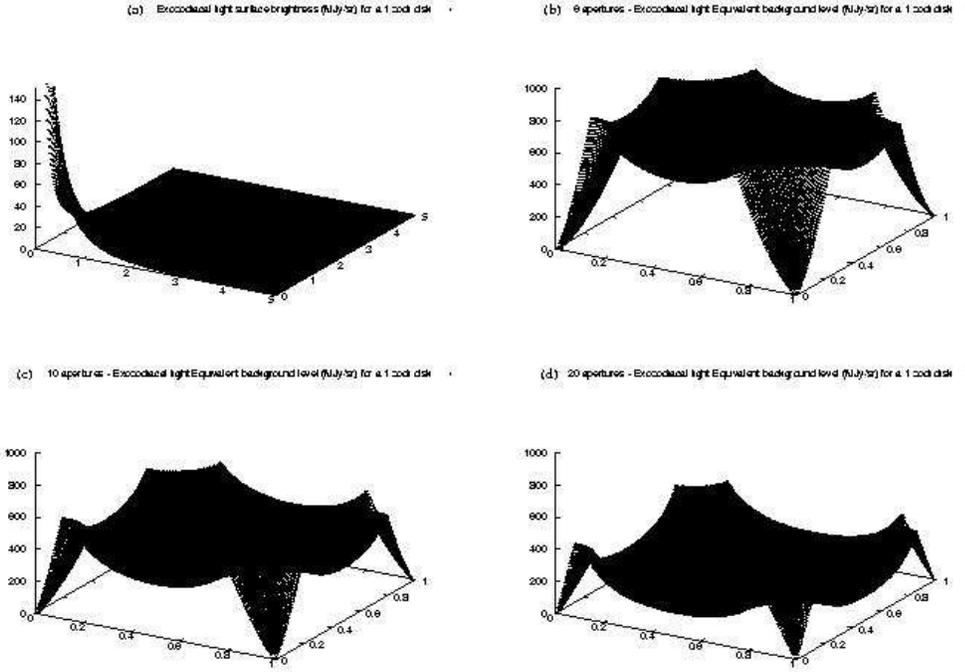}
\caption{(a) True exozodiacal light distribution in a 1 Zodi system at $10\:pc$ (spatial coordinates are in AU). (b), (c) and (d) Exozodiacal light Equivalent background level when this system is observed respectively with a 6, 10 and 20 apertures interferometer (spatial coordinates in units of $\delta$). In (b), (c) and (d), the effect of the nulling coronagraph has been simulated by multiplying the Exozodiacal light Equivalent background level by the transmission profile of the phase mask coronagraph.}
\end{figure}

\vspace*{5in}

$B_{EZ}$ is a $\delta$-periodic function of $x$ and $y$ and is shown (for $0<x<\delta$ and $0<y<\delta$) in Fig. B.1 for $N=6$, $N=10$ and $N=20$. As a comparison, the true exozodiacal cloud surface brightness is also shown in the same figure. 

This figure illustrates the effect of having a sparse aperture: the relatively small angular spacing between diffraction peaks (or fringes in 1 dimension) results in a ``mixing'' of a large amount of exozodiacal light into the light of the companion. In this figure, $B_{EZ}$ was computed for a 60m baseline interferometer at $10 \mu m$ observing a 1 Zodi cloud at $20\:pc$. $B_{EZ}$ is independent of the individual sub-apertures diameter.
With 6 sub-apertures, the median Exozodiacal light Equivalent background level across the field is $603\: MJy.sr^{-1}$ and $\delta$ is 0.8 AU ($355\: MJy.sr^{-1}$ and 1.1 AU for 10 apertures, $173\: MJy.sr^{-1}$ and 1.5 AU for 20 apertures). By replacing $B_{EZ}$ by this value in equation B.19, we obtain (for 6 apertures, $3\:m$ diameter each, $60\:m$ baseline):
\begin{equation}
N_{EZ} = 60.5\: ph.s^{-1}.\mu m^{-1}.
\end{equation}

{\bf Signal to noise ratio}\\
The signal to noise ratio for the detection of a companion is (noiseless detector)
\begin{equation}
\frac{S}{N} = \sqrt{T_{eff}} \frac{N_{pl}}{\sqrt{N_{pl}+N_{st1}+N_{st2}+N_{Z}+N_{EZ}}},
\end{equation}
where $T_{eff}$ is the effective exposure time (in $s.\mu m$).
With the values computed above, we obtain a S/N of 3 in $101\: s.\mu m$ effective exposure time, corresponding to a 17mn exposure time with a $0.1$ efficiency (transmission of optics multiplied by the quantum efficiency of the detector) and a $1 \mu m$ bandwidth.
The residual star light and the zodiacal light are by far the most important sources of noise in this computation.

{\bf Non-redundant arrays}\\
When using a non-redundant array, the PSF of the companion is less contrasted: its light is spread over a larger area. With a redundant array, most of the light (light inside the diffraction peaks) occupies a fraction $F_{red}$ of the PSF, with
\begin{equation}
F_{red} = \frac{1}{N}.
\end{equation}
For a non-redundant array, it is difficult to estimate which diffraction peaks should be considered as part of the the area in which the companion's signal should be taken into account for the S/N computation. Rather than measuring the signal in discrete domains, as was done for the redundant array above, one should deconvolve the snapshot image to recover the companion. However, such a deconvolution is not easy to achieve because of the position-dependency of the PSF introduced by the use of a coronagraph, and is beyond the scope of this work.
For this simple estimate, we model the non-redundancy of the array as an increase of the relative area $F$ of the diffraction peaks in the PSF and a decrease of the flux they contain.
The number of star leakage, zodiacal and exozodiacal photons ``mixed'' with the companion's photon is then increased by $\frac{F}{F_{red}}$ relative to the redundant array configuration. With a non-redundant array such as the one shown in Fig. 1, most of the companion's photons are contained in half of the total PSF area: in the PSF of the non-redundant 6-apertures array, 80\% of the flux is contained in 50\% of the PSF's area. Therefore, with $F=0.5$ we expect the noise term of equation (28) is to be multiplied by 
\begin{equation}
\sqrt{\frac{F}{F_{red}}} = \sqrt{\frac{N}{2}},
\end{equation}
which is equal to $\sqrt{3}$ in our example, and the signal (planet's photons) is multiplied by $E=0.8$. Note that the photon noise from the companion itself is not increased, but since it represents a very small fraction of the noise, we can ignore it in this simple estimate.
This simple estimate shows that using a non-redundant array will increase the exposure time required for detection of the companion by approximately $\frac{N}{2 \times {E}^2}$, which in our example is 4.7.
With a larger number of apertures, the difference of S/N becomes larger between a redundant and a non-redundant array, but it also becomes easier to obtain a good (u,v) plane coverage without using a very non-redundant array.

\end{document}